Fig. 1

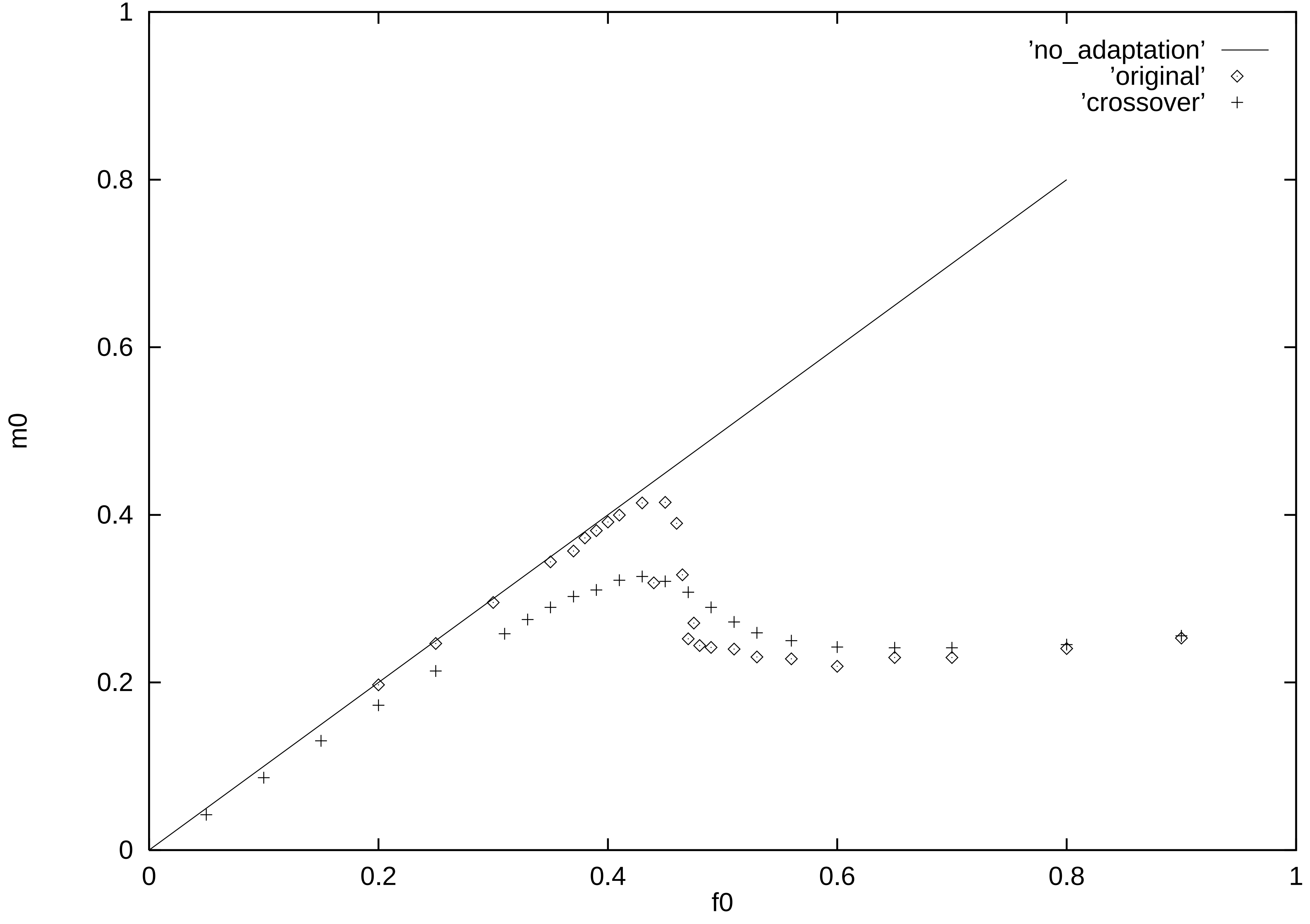

Fig. 2

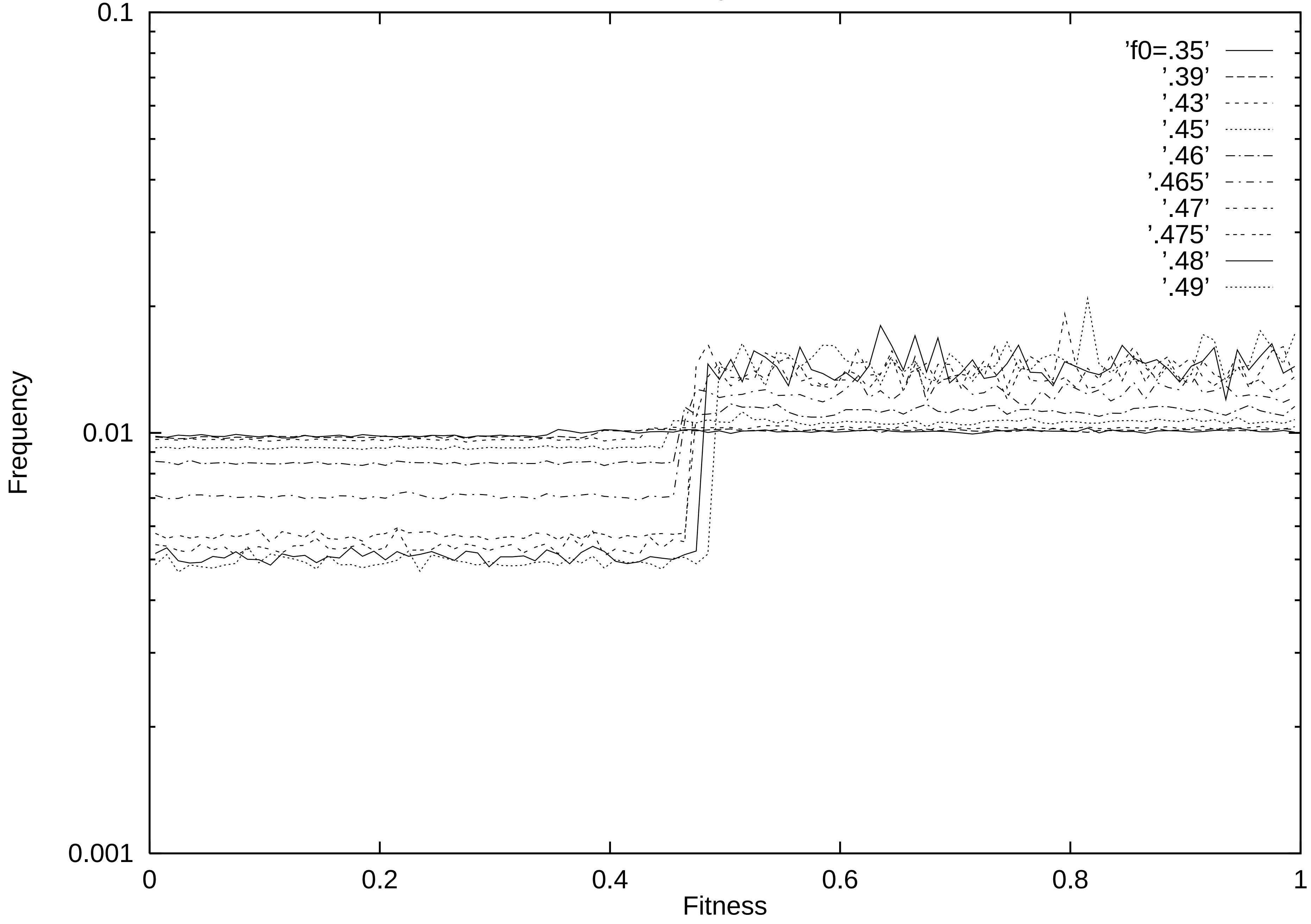

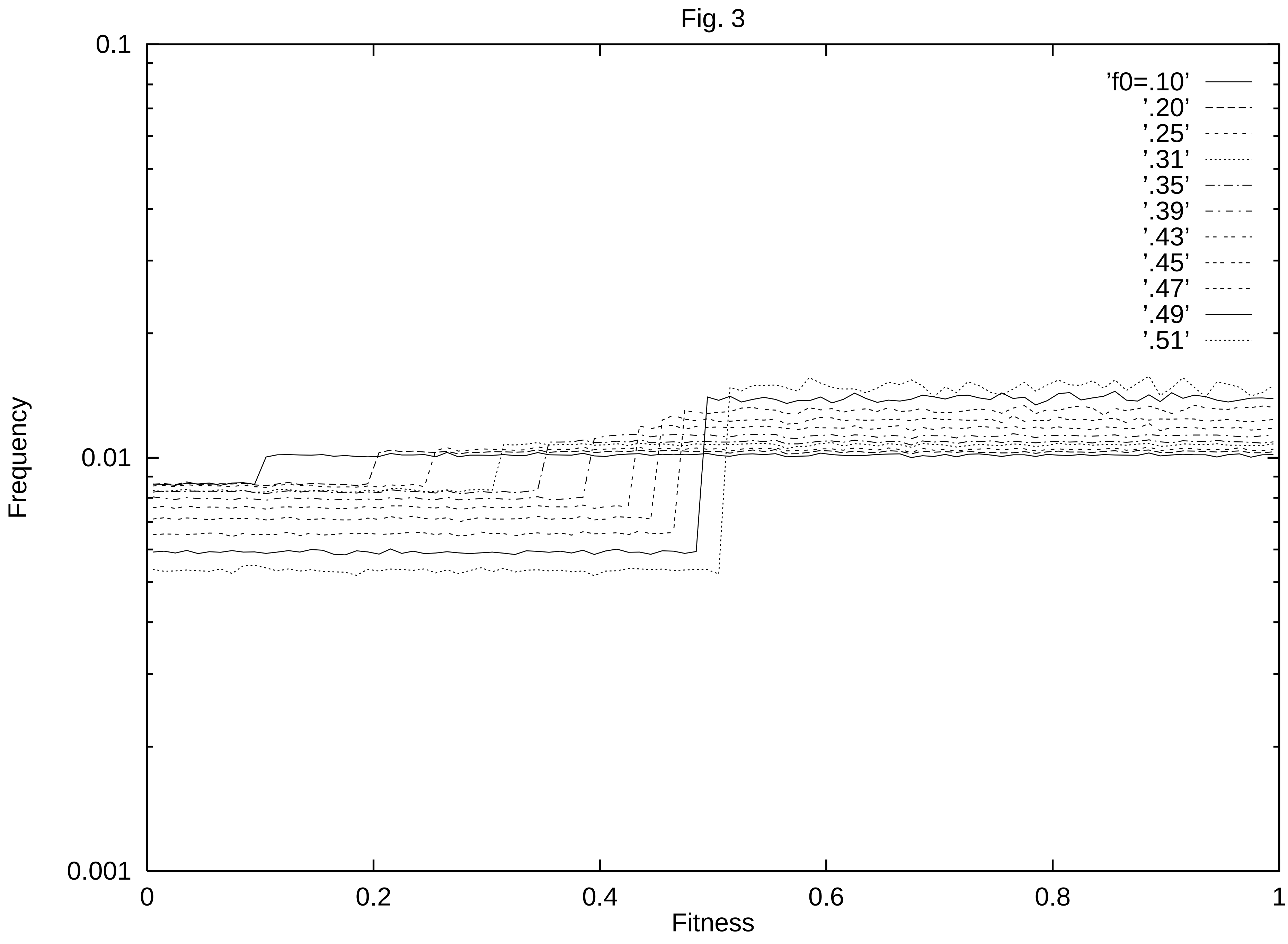

Fig. 4

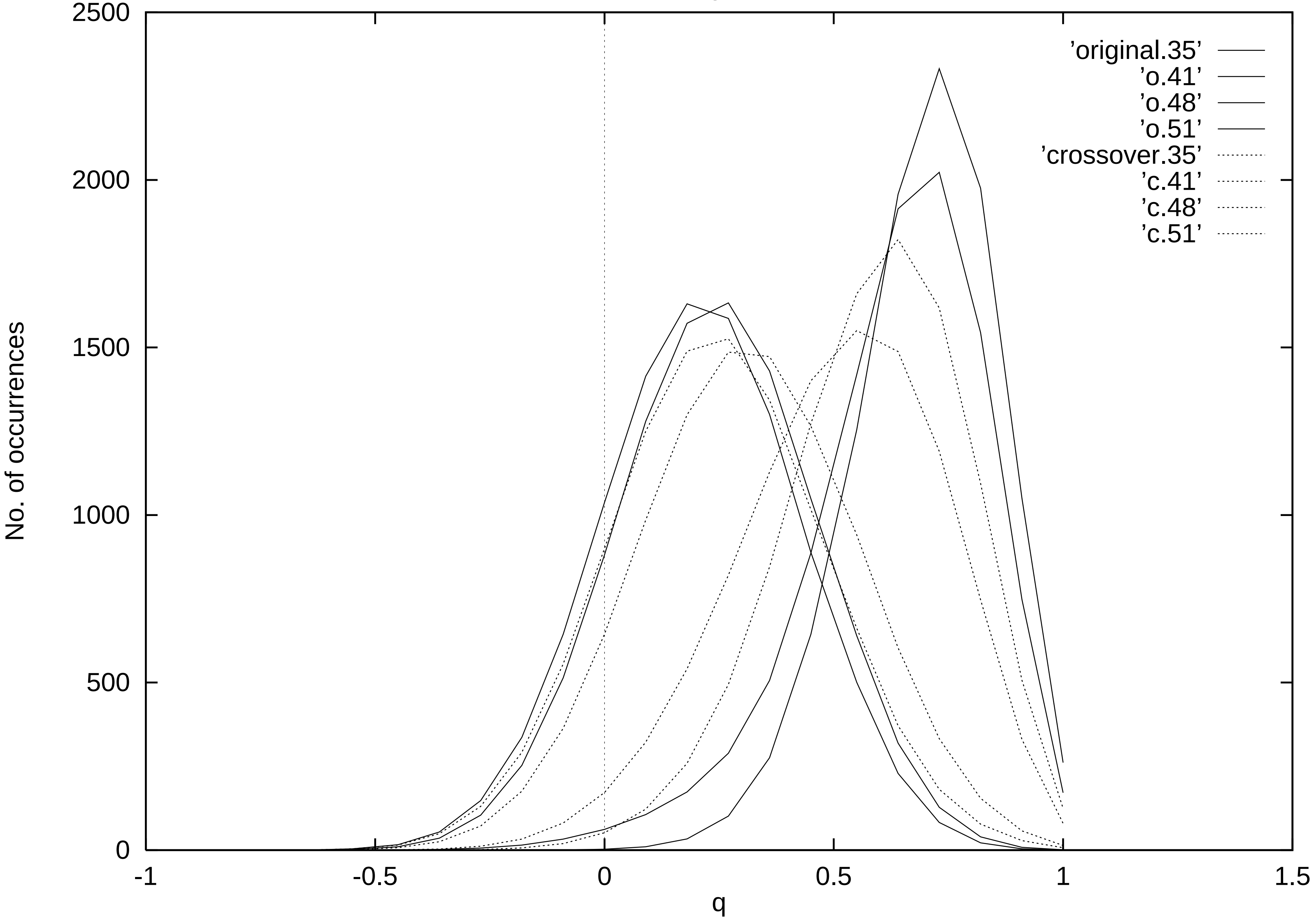

Fig. 5

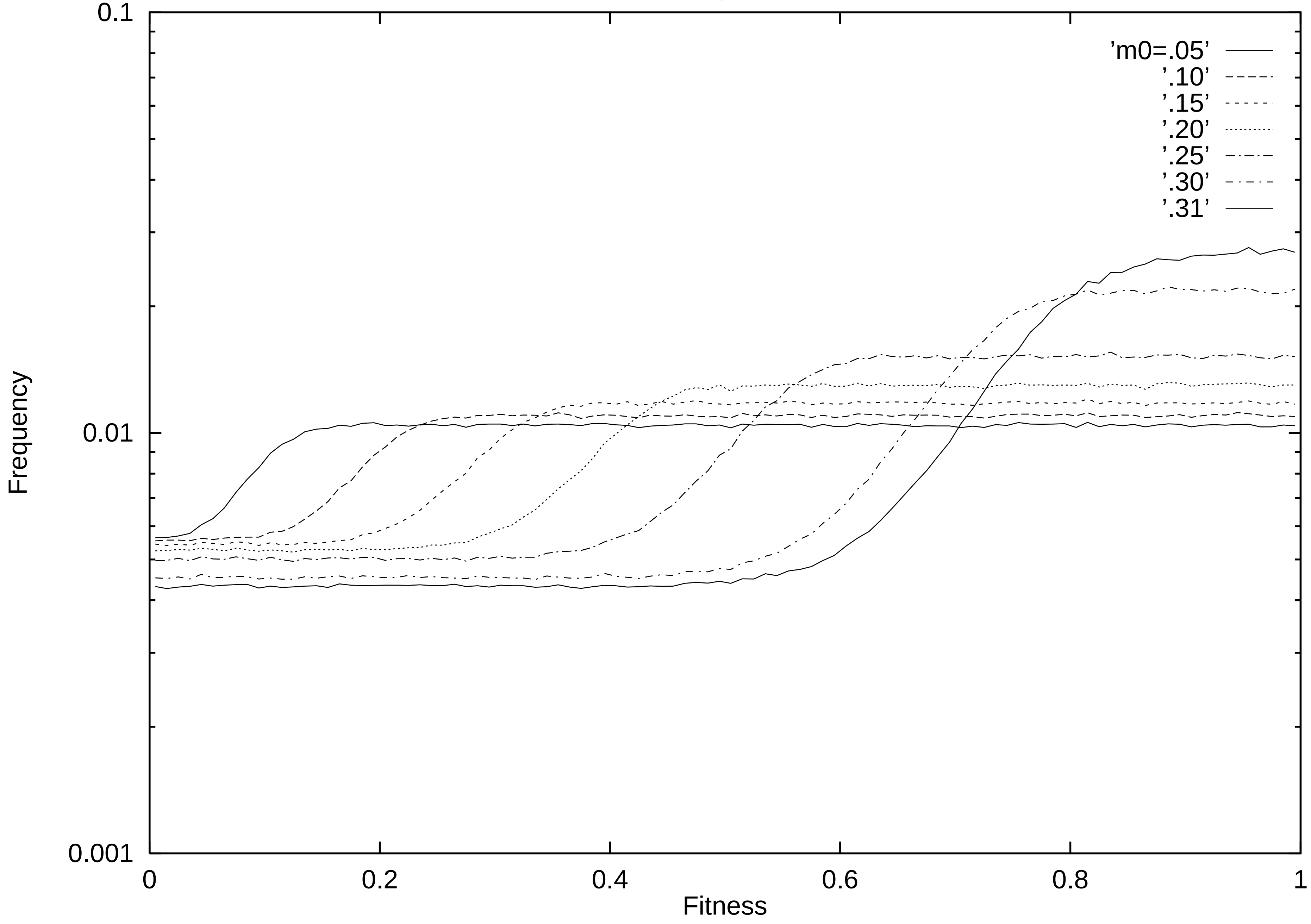

# Self-Organization in Models of Populations with Two-Parent Reproduction on a Rugged Fitness Landscape

Jan Finjord

*Stavanger College, Stavanger, Norway**

*and*

*Center for Nonlinear Studies*

*Los Alamos National Laboratory*

*Los Alamos, NM, USA*

(May 19, 1995; email: finjord@hsr.no)

**Abstract.** The two-parent reproduction model of Derrida and Peliti is simulated on a rugged fitness landscape. Fixed fitness values for each possible genotype are assigned randomly, with all fit individuals having the same probability of reproduction. The previously observed transition to a self-organized state of the population with less recombinational load, implies an abrupt change of genetic overlap distributions, showing up characteristics of a phase transition. A crossover variant of the model has a smoother transition to the adapted regime, with a residual collective adaptation for small values of the threshold in fitness. When a geographical constraint (shortest possible distance) on pairwise reproduction in a population arranged one-dimensionally is imposed, a poised state results, suggestive of self-organized criticality.

## I. INTRODUCTION

Some models proposed for molecular evolution have analogies in physics. Kauffman's *NK* model [1] for evolution on rugged fitness landscapes is identical to the random-energy spin-glass model [2] in the limit $K = N - 1$. The *NK* model is related to another model for molecular evolution which is also a spin-glass analogy [3]. In another line of development, a model proposed by Derrida and Peliti for evolution on a flat fitness landscape [4] is analogous to the annealed random map model [5]. Drawing on such analogies, properties and solutions of the evolution models could be found.

Further developments of the models have introduced traits which to some extent have let go the analogy with disordered systems. Interest has centered on general properties of the evolution models as such. In the Derrida-Peliti model, traits like two-parent recombination [6], genetic overlap restrictions [7], and a rugged fitness landscape [8] have been introduced. The latter differs from the *NK* landscape in that there are no epistatic interactions; each possible genome has a fixed fitness, which has been assigned randomly. For a recent review, see [9].

Analogies with physics would be re-established on another level if the models could be shown to possess properties like critical transitions, self-organization [1], or even self-organized criticality (SOC) [10]. Indeed, a descendant of the Derrida-Peliti model has been shown [11] to possess an adaptation property reminiscent of the first two. On the other hand, it has been claimed [12] that the *NK* model probably does not possess SOC.

The present paper addresses the two-parent version of the Derrida-Peliti model on a rugged fitness landscape. The nature of the observed adaptation is studied, and a search for SOC in the model and variants of it is reported on.

## II. MODEL

### A. Reproduction mechanisms

We consider a population made up of a fixed number, $M$, of individuals whose genome is characterized by $N$ binary units $S_i^\alpha$, with

$$S_i^\alpha \in \{-1, +1\}, \qquad 1 \le \alpha \le M, \qquad 1 \le i \le N \tag{1}$$

*Permanent address

In the original model with haploid reproduction [4], a new generation with $M$ members is born from the one at time $t$ with each new individual equal to its only parent $G(\alpha)$, except for mutations with a bare rate $\mu$ per genome unit:

$$S_i^\alpha(t+1) = \epsilon_i^\alpha(t)\, S_i^{G(\alpha)}(t), \qquad \epsilon_i^\alpha \in \{-1,+1\}, \qquad \overline{\epsilon_i^\alpha} = e^{-2\mu} \tag{2}$$

The random variable $\epsilon_i^\alpha$ thus takes the values $\pm 1$ with probability $\frac{1}{2}(1 \pm e^{-2\mu})$. An extension of this model to two-parent reproduction was introduced subsequently [6], with two parents $G^{(1)}(\alpha)$ and $G^{(2)}(\alpha)$ of each new individual $\alpha$, chosen at random in the population:

$$S_i^\alpha(t+1) = \epsilon_i^\alpha(t) \left[\xi_i^\alpha(t) S_i^{G^{(1)}(\alpha)}(t) + (1-\xi_i^\alpha(t)) S_i^{G^{(2)}(\alpha)}(t)\right], \qquad \xi_i^\alpha \in \{0,1\}, \qquad \overline{\xi_i^\alpha} = \frac{1}{2} \tag{3}$$

The random variable $\xi_i^\alpha$ determines from which one of the chosen parents the $i$'th genome unit will be inherited.

The two reproduction mechanisms were studied initially with a flat fitness landscape. In analogy with the theory of disordered systems, the necessity of introducing both a population average and a time average was stressed. The haploid version was shown [4] to possess most of the features of mean field spin glasses, being equivalent to the annealed random map model [5]. The two-parent version was shown to have the same average properties as the haploid one, but with diffent characteristics of fluctuations in genetic distance: In the infinite-population limit the fluctuations disappear, and most of the analogy with disordered systems seems irrelevant in the two-parent case [6]. Analytical expressions for average quantities were derived in both cases.

The two-parent model described above was introduced as an analogy to sexual reproduction in biology (without two sexes). To get an indication of the influence of the particular mechanism of two-parent inheritance on the results, in this paper we will also introduce a crossover model:

$$S_i^\alpha(t+1) = \epsilon_i^\alpha(t) \left[H(N\rho^\alpha(t) - i) S_i^{G^{(1)}(\alpha)}(t) + (1 - H(N\rho^\alpha(t) - i)) S_i^{G^{(2)}(\alpha)}(t)\right], \qquad \rho^\alpha \in <0,1> \tag{4}$$

The value of the Heaviside function $H$ determines the crossover point, the position of which is given by the random variable $\rho^\alpha$. With a finite genome length $N$, the specification is modified slightly so that each parent will pass on at least one genome unit, with each crossover point equally probable.

### B. Restrictions on reproduction by genetic overlap, fitness, and geography

One measure of the genetic structure of the population is given by the distribution of the genetic overlap $q^{\alpha\beta}$ between individuals $\alpha$ and $\beta$, which is related to their Hamming distance $\nu^{\alpha\beta}$:

$$q^{\alpha\beta} = \frac{1}{N}\sum_{i=1}^{N} S_i^\alpha S_i^\beta \tag{5}$$

$$\nu^{\alpha\beta} = \frac{1}{4}\sum_{i=1}^{N}(S_i^\alpha - S_i^\beta)^2 = \frac{N}{2}(1 - q^{\alpha\beta}) \tag{6}$$

In the present paper, no restrictions will be put on the allowed values of $q^{\alpha\beta}$ for a pair of prospective parents; we will consider the distribution of $q^{\alpha\beta}$ under the influence of fitness constraints on evolution. In [7,13], however, the results of imposing a lower limit $q_0$ for the allowed overlap values in two-parent evolution on a flat fitness landscape have been studied in the infinite-genome limit. The nonlinearity introduced by the cutoff in $q^{\alpha\beta}$ prevents analytical solutions. Qualitatively, an ultrametric overlap distribution and fluctuations similar to those of the haploid model was observed, inspiring a notion of species formation in the model. A recent paper [14] considers such a lower fecundity limit in a model with discrete geographical locations, and conjectures that the time scales emerging in the resulting speciation can be put in correspondence with palaeontological data.

Introduction of natural selection can be made by ascribing fixed fitness values randomly to each of the $2^N$ possible genomes, and introducing a lower allowed fitness value $f_0$ or a constant fraction $m_0$ of the population with the lowest fitness values as unfit for reproduction. For evolution on a rugged fitness landscape of this type, with $f_0$ specified, and all fit individuals having the same *a priori* probability of reproduction, the two-parent reproduction mechanism of eq. (3) shows [11] properties markedly different from the haploid one [8]: Depending on the values of $f_0$ and $\mu$, in the two-parent case evolution can create a new regime of collective adaption with a higher frequency of fit genotypes, thus introducing a notion of recombinational fitness between pairs of individuals (a correlated variable) on an uncorrelated fitness landscape. One purpose of the present paper is to investigate further the properties of the two regimes and the transition between them.

Geographical dependence can be introduced by arranging the individuals in one or more dimensions and imposing limits on the geographical distance between parents. In the simulations underlying this paper the one-dimensional case with periodic boundary conditions, with parents located as close to each other and to the new individual as possible, was considered as an alternative to random choice of parents.

The population description in eq. (1) can also be used for an introduction of a genome structure in the species evolution model of Bak and Sneppen [12,15,16]. By letting a genome define a species, ascribing a fixed fitness value to each possible genome as described above, and describing evolution as replacing the actually least fit species and a fixed number of others per time step, an equivalent to the Bak-Sneppen model is obtained. In the simulations underlying the present paper, the version with replacement of the least fit one and its nearest neighbor on each side was considered.

## III. SIMULATIONS

### A. Parameter specifications and implementation details

In our simulations, the parameter values mostly used are

$$M = 100, \qquad N = 22, \qquad \mu = 0.015 \tag{7}$$

Comparisons for some other $\mu$ values were also made. Process averages were taken over 260000 time steps following initial relaxation over 60000 steps; in a region of slow convergence for the original model, 5 times as many steps were included in the average. For the genetic overlap distributions, 20000 steps following 20000, respectively, were used. Bak-Sneppen results were obtained with $10^6$ time steps, following 50000 initial steps, for a species collection of size $M = 250$.

Fixed fitness values for the $2^N$ genotypes were assigned randomly over the interval $[0, 1]$. The fraction of unfit genotypes could then effectively be identified with the lower allowed fitness value $f_0$. The `rand` generator [17] of a Sun SPARCstation was used in the Monte Carlo simulations, including the assignment of fitness. As a guard against spurious effects of one particular choice of generator, checks with fitness assignment using `RAN2` or alternatively `RAN3` [18] were made. Initial genomes were assigned randomly; checks confirmed that assigning a common random initial genome to all individuals instead, made no difference.

In the case with individuals arranged one-dimensionally, the parents of an individual in a given location were chosen as the preceding individual in that location and its nearest neighbor to a randomly chosen side. If the preceding individual or its neighbor were not fit, successive random nearest-neighbor searches for fit individuals were made, starting from the new individual's location, independently for each parent.

### B. Two-parent reproduction, given lower fitness limit

Two-parent reproduction was investigated with both the reproduction prescription in eq. (3) [6] and the crossover version in eq. (4). Fig. 1 shows the fraction of actually unfit individuals, $m_0$, plotted against the fraction of unfit genotypes, $f_0$, with the value of the latter as a constraint. With the original model, the drop found by Peliti and Bastolla [11] is reproduced using the original model, whereas the crossover model shows a markedly smoother transition in the same region of $f_0$. For higher $f_0$ values, the $m_0$ values for both mechanisms seem to approach a common limit. For $f_0$ values below the threshold, however, $m_0$ values with eq. (4) fall consistently below those with eq. (3). Linear asymptotic dependences of $m_0$ on $f_0$ are observed, with asymptotic slope value 1 for the original model and a parameter-dependent value for the crossover model. Shorter runs for some other $\mu$ values showed the same qualitative behavior, with the position of the transition region as well as $m_0$'s asymptotic value roughly proportional to $f_0$.

Figs. 2 and 3 shows the actual distribution of fitness values after reproduction, with eqs. (3) and (4) used, respectively, and $f_0$ as a parameter. The qualitative difference observed below threshold in fig. 1, present itself in the crossover model as a decreased tendency to produce offspring with fitness value below $f_0$.

Fig. 4 shows distributions of $q^{\alpha\beta}$ with eqs. (3) and (4), for four $f_0$ values close to the adaptation threshold. Another aspect of the collective adaptation shows up: As $f_0$ increases through the threshold, the peak in the overlap distribution moves from a low value to one close to 1. In comparison, except in the crossover region the difference in distribution between the original model and the crossover model is small, although distinct. The crossover mechanism provides the more gradual transition to the collective adaptation regime, since some adaptation is present there for $f_0$ values where adaptation has not yet occurred in the original model.

### C. Two-parent reproduction, given fraction of unfit individuals

As a number constraint, the $m_0 M$ least fit individuals were considered unfit for reproduction. With the parameter values used, the fitness distributions became unstable and dissolved into a few spikes above the $m_0$ values 0.33 and 0.28 for eqs. (3) and (4), respectively. For smaller $m_0$, the distributions for eq. (3) were flat within plotting accuracy (cf. fig. 2), while the curves for eq. (4) had rounded thresholds instead of the discontinuous jumps in fig. 3. The instability results from the double-valuedness in $m_0$ shown in fig. 1, where large enough values can correspond to two different regimes of adaptation.

### D. Nearest-neighbor reproduction

With a given lower fitness limit, fitness distributions like those in fig. 3 for random choice of parents are obtained. The new traits are that this happens also for the original model (eq. 3), although with smaller jumps than for the crossover model; and that the transition to the regime of collective adaptation is even smoother than for the crossover model with random choice of parents.

Also with a given number of unfit individuals, both models show up a rounded threshold in the fitness distributions. Fig. 5 shows the crossover case. An abrupt transition to a spiky fitness distribution again takes place when $m_0$ exceeds a certain value. New is that as $m_0$ increases towards that value, the threshold gets a proportionally larger displacement to the right, until it literally hits the right frame edge and the transition to a spiky distribution occurs. A smaller but still distinct displacement was observed when the original two-parent model (with eq. (3)) was used.

### E. Species evolution model

The minimal introduction of a genome structure in the Bak-Sneppen model, resulted in an SOC state with the same threshold position as in ref. [12], within the error bars cited there. Above threshold, the oscillations in fitness frequency were of order 20 % for the parameters used, comparable to the accuracy of ref. [12] considering that the analogy of $M = 4096$ was used there. Runs were also made with the species collection changed according to the Bak-Sneppen prescription first in each step, followed by nearest-neighbor recombination. No poised state corresponding to the one in ref. [12] was then obtained; the resulting fitness distribution was much like the one found in section III C for $m_0 = 0.03$.

## IV. DISCUSSION

For two-parent reproduction with a given lower fitness limit, fluctuations in the $m_0 = m_0(f_0)$ curve for eq. (3) close to the threshold in $f_0$ were observed, with a considerable dependence on the generator initialization values. The magnitude of such possible fluctuations is shown by including the 'stray' value for $f_0 = 0.44$ in the figure (with another generator initialization value, the point moved upwards, giving a smooth curve). The irregularities in the corresponding curve in fig. 1 of ref. [11] on both sides of the threshold can possibly be ascribed to such effects. They may be interpreted as fluctuations between the two different types of overlap distributions, caused by the finite $M$ and $N$ values. The results for the crossover case, eq. (4), showed less fluctuations. The self-organization property occurring as a collective adaptation results in flat process-averaged distributions in fitness on both sides of the $f_0$ value, with a discontinuous jump in between. On the basis of the abrupt change of $q^{\alpha\beta}$ distributions for the original model, and the fluctuation property, we hypothesize that a dynamical phase transition occurs as $f_0$ increases through a threshold. Whereas the $M$ value used in the simulations is slightly larger than that of ref. [11], the $N$ value is only about one fourth. This may explain the apparently sharper transition observed in ref. [11], compared to our results using eq. (3). Further consideration of the nature of the transitions—unless a quantitative analytical theory can be found—will have to await better statistics, obtained for larger $M$ and $N$ values. With CPU times for each curve in fig. 2 of this paper typically of order 10 hours on a SPARCstation, and still limited statistics, this appears like a task for massively parallel computation.

Neither the original mechanism nor its crossover variant should necessarily be considered close approximations to the replication strategy of terrestrial carbon-based life. Rather, they are candidates for optimization tools in generalized populations. We note that the recombinatorial load [19] would be most efficiently reduced with a crossover type strategy, for small values of the lower fitness limit $f_0$.

In ref. [7] where the limit $N \to \infty$ was used, the $q^{\alpha\beta}$ distributions appeared as consisting of one or more sharp spikes. With finite and low $M$ and $N$ values as in the present case, the distributions show up broad peaks, and $q^{\alpha\beta}$ is no longer positive definite.

The simulations described in sections III C, III D and III E searched for actual displacements of the threshold in fitness of the type described in ref. [12]. Such an effect was found only for nearest-neighbor reproduction in one dimension with a given number of unfit individuals. (The other observed presence, in the Bak-Sneppen model with a genome, should be considered a trivial confirmation of previous results.) The threshold position was observed to be dependent on the values of the parameters in eq. (7). In itself that is not an argument against SOC; in ref. [15] the threshold position was shown to depend on the number of changed species, and f. i. $\mu$ has a similar function in the present model, since it measures the amount of random input into the population per generation.

It therefore appears that at least in one dimension, geographical limitations on reproduction holds the possibility of inducing a poised state. In fig. 5 this is seen as a translation of the starting point of the rise in the curve (for not too small $m_0$) to a fitness value considerably above $m_0$. Also the original model shows this effect; the crossover model was chosen for the figure for maximum clarity. A poised state is one of the characteristics usually present in SOC. If further study should confirm also the presence of power laws, that would seem to be the first instance of SOC in evolution models with a genome introduced on the individual level, and in particular for two-parent reproduction.

The suggestion of ref. [15] to represent each species by a population of individuals reproducing sexually, could conceivably hold the possibility for another occurrence of SOC. However, with a large collection of species, each consisting of a large number of individuals, the undertaking of such calculations would necessitate the use of parallel computers.

## V. CONCLUSION

The transition to a collectively adapted state in the original two-parent Derrida-Peliti model is characterized by an abrupt change in genetic overlap distributions, suggesting a critical transition. A crossover version of the model scans genome space more efficiently, having a residual adaptation for low values of the fitness threshold and the same behavior as the original version for high values of the threshold, with a smooth transition in between. Finally, an example *in silico* [20] has been presented where introduction of a geographical constraint on reproduction in the two-parent model results in a poised state, which suggests the presence of self-organized criticality.

The author thanks L. Peliti for providing a copy of ref. [14] prior to publication and for being made aware of ref. [11], and S. Kauffman and B. Levitan for a discussion of the results. It is also a pleasure to thank G. Doolen for the invitation to LANL and for providing computation equipment, and S. M. Skjæveland for allocating the project support which made the stay possible. The Norwegian Research Council and Stavanger College also provided support, the latter in particular by granting sabbatical leave.

---

## FIGURE CAPTIONS

FIG. 1. Fraction of unfit individuals as a function of the fraction of unfit genotypes. Diamonds, original model; crosses, crossover model.

FIG. 2. Fitness distributions in the population for the original model, with $f_0$ as a parameter.

FIG. 3. Fitness distributions in the population for the crossover model, with $f_0$ as a parameter.

FIG. 4. Distributions of genetic overlaps $q_{\alpha\beta}$ in the original model (full lines) and the crossover model (dotted lines). For both models, the curves correspond to $f_0 = 0.35, 0.41, 0.48,$ and $0.51$, respectively, from left to right.

FIG. 5. Fitness distributions for the crossover model with nearest-neighbor reproduction and a given fraction $m_0$ of unfit individuals, with $m_0$ as a parameter. From $m_0 H = 32$ on, the distribution changed into a collection of spikes.